# Multiple-view spectrally resolved x-ray imaging observations of polar-direct-drive implosions on OMEGA

R. C. Mancini, H. M. Johns[a], T. Joshi, D. Mayes and T. Nagayama[b]
Physics Department, University of Nevada, Reno, NV

S. C. Hsu, J. A. Baumgaertel, J. Cobble, N. S. Krasheninnikova, P. A. Bradley, P. Hakel,
T. J. Murphy, M. J. Schmitt, R. C. Shah, I. L. Tregillis, and F. J. Wysocki
Los Alamos National Laboratory, Los Alamos, NM

**Abstract**

We present spatially, temporally, and spectrally resolved narrow- and broad-band x-ray images of polar-direct-drive (PDD) implosions on OMEGA. These self-emission images were obtained during the deceleration phase and bang time using several multiple monochromatic x-ray imaging instruments fielded along two or three quasi-orthogonal lines-of-sight including equatorial and polar views. The instruments recorded images based on K-shell lines from a titanium tracer located in the shell as well as continuum emission. These observations constitute the first such data obtained for PDD implosions. The image data show features attributed to zero-order hydrodynamics. Equatorial view synthetic images obtained from post-processing a 2D hydrodynamic simulation are consistent with the experimental observation. Polar view images show a pentagonal pattern that correlates with the PDD laser illumination used on OMEGA, thus revealing a 3D aspect of these experiments not previously observed.

1. Introduction

Polar-direct-drive (PDD) implosions have attracted significant attention in the last decade because they provide an opportunity to perform direct-drive implosion experiments at the National Ignition Facility (NIF) [1] using the current configuration of NIF laser beams, i.e. a configuration customized for indirect-drive implosions [2]. This direct-drive approach extends the application of NIF to investigate alternative inertial confinement fusion (ICF) schemes. Extensive numerical simulation work has been done to design targets and model PDD implosions for both OMEGA [3] and NIF [4-14]. However, the majority of the experimental work so far has been done at OMEGA, with few experiments at NIF [4-6, 14-17].

In the process of developing and characterizing a PDD ICF experimental platform intended to study mix in the presence of thermonuclear burn we have obtained spatially, temporally, and spectrally resolved x-ray images of K-shell emission from titanium shell-dopant material mixed into the core for PDD implosions on the OMEGA laser facility [15-17]. The primary motivation for these x-ray imaging measurements was to demonstrate the ability to obtain spatially resolved images of shell/fuel mix, which, in combination with neutron images in future higher-yield shots on the NIF, could potentially provide experimental constraints on models of thermonuclear burn in the presence of mix [18]. Coincidentally, our results also provide basic data that could be useful in the context of validating PDD hydrodynamic simulations as well as evaluating and optimizing PDD platforms aimed at ignition on NIF [15]. Although this work was not motivated by PDD ignition, we did benefit from the extensive prior work done on finding optimum laser pointing for PDD on OMEGA [6].

Most of the PDD imaging data from OMEGA shots have focused on diagnosing ablator symmetry in the early phase of the implosion via radiography [2,4-6,14,16]. Very little has been done on imaging the self-emission of the implosion core in the deceleration phase and near peak compression in order to understand the details of the hot dense plasma assembly at the collapse of the implosion. In addition, all of the prior PDD imaging data has been obtained only along a quasi-equatorial line-of-sight (LOS), never from along a polar LOS. In this work, we present the first PDD imaging data that is spatially, spectrally, and temporally resolved, along *both* the equatorial and polar LOS. This was accomplished by fielding two or three identical x-ray multi-monochromatic imager (MMI) instruments [19-22] in quasi-orthogonal ports (i.e., TIM: ten-inch-manipulators) on the OMEGA target chamber. The best data showing K-shell emission from the titanium shell tracer came from late in the implosion (well after the laser drive was turned off) and thus provide information on the shape of the implosion core during the deceleration phase, as well as electron temperature and density after spectral data unfolding, in the few hundred picoseconds leading up to bang time. The key observations reported here based on MMI self-emission images are: (1) systematic differences in the images between equatorial and polar LOS, (2) polar views reveal a consistent pentagonal, petal pattern that correlates with the PDD laser illumination pattern used on OMEGA and can be interpreted as a laser imprint effect, and (3) the equatorial views reveal a "hamburger" or "double-bun" spatial structure. All three observations reveal zero-order features in that they are observed in both narrow- (i.e., at a K-shell titanium line energy) and broad-band MMI images, i.e. they are independent of the x-ray photon energy range of the observation. This paper is organized as follows. Section II describes the details of the experiment. Section III provides the experimental results and discussion. Section IV provides conclusions and a summary.

2. **Experiments**

The experiments were performed at the OMEGA Laser Facility [3] using the 40-beam configuration for PDD implosions [6], and were part of the defect induced mix experiment (DIME) campaign of Los Alamos National Laboratory [15]. The 40 laser beams were arranged in 6 rings, 3 of which illuminated the northern hemisphere of the spherical shell and the other 3 the southern hemisphere. In each group of 3 rings, there were 5, 5 and 10 laser beams, respectively. Table 1 shows the beam pointing parameters of each ring in terms of the polar angle $\theta$ of the cone and the displacement Δr [6]. This displacement represents the offset from target center of the laser beam measured along a direction perpendicular to the beam axis.

Table 1: laser beam pointing parameters of the northern hemisphere 20 beams for polar drive shots at OMEGA. A complementary set of parameters is used for the group of 20 beams that illuminate the southern hemisphere [6].

|  | Ring 1, 5 beams | Ring 2, 5 beams | Ring 3, 10 beams |
| --- | --- | --- | --- |
| θ (degrees) | 21.4 | 42.0 | 58.8 |
| Δr (μm) | 90.0 | 150.0 | 150.0 |

Beam smoothing included the distributed phase plates (DPP), distributed polarization rotators (DPR) and smoothing by spectral dispersion (SSD) options available at the OMEGA laser facility resulting in a beam-to-beam variation of about 3% root-mean-square [3]. All laser beams were configured to have the same energy, and delivered a total of about 17.5kJ of UV laser energy on target with a 1ns square pulse shape. Targets consisted of spherical plastic shells that had an outer diameter of about 850μm, wall thicknesses in the 16μm to 19μm range, and an outer aluminum coating of 0.1μm for sealing purposes. They were filled with 5atm of deuterium gas. A titanium-doped (1.8% atomic concentration) tracer layer 2μm thick was added to the shell either on the inner surface of the shell or embedded in the shell at a given distance from the inner surface for spectroscopy diagnosis purposes. Specific target details for the shots discussed in this paper are shown in Table 2.

Table 2: target outer diameter (OD), wall thickness (ΔR), depth of Ti tracer layer relative to shell's inner surface, deuterium filling pressure ($P_{D2}$), UV laser energy on target ($E_L$), and neutron yield (n). All shots were PDD but shot 65036 that was driven with a 60-beam symmetric drive.

| Shot | OD (μm) | ΔR (μm) | Ti layer depth (μm) | $P_{D2}$ (atm) | $E_L$ (kJ) | n |
| --- | --- | --- | --- | --- | --- | --- |
| 60893 | 871 | 15.7 | 0.0 | 5.7 | 17.5 | $1.1 \times 10^{10}$ |
| 65036 | 839 | 18.7 | 0.0 | 5.0 | 22.8 | $3.8 \times 10^{10}$ |
| 67066 | 849 | 19.0 | 0.0 | 5.2 | 17.7 | $5.1 \times 10^{9}$ |
| 67069 | 849 | 19.0 | 0.0 | 5.0 | 17.5 | $5.5 \times 10^{9}$ |
| 67071 | 877 | 19.0 | 1.3 | 5.1 | 17.7 | $1.1 \times 10^{10}$ |
| 67072 | 868 | 19.0 | 0.7 | 5.1 | 17.6 | $8.2 \times 10^{9}$ |

Gated, spectrally resolved x-ray images were recorded with the MMI instruments available at OMEGA for fielding in direct-drive implosions [19-23]. The MMI imagers viewed the implosion along both equatorial and polar LOS. The MMI instrument consists of a pinhole array, a multi-layered mirror (MLM), and a micro-channel-plate (MCP) framing camera detector. In particular, the framing cameras were equipped with 4 MCP's thus providing the opportunity for recording up to 4 snapshots of the implosion. The pinhole array had a few hundred pinholes, each 10μm in diameter, laser-drilled on a 50μm thick tantalum substrate, and hexagonally packed with an average separation of 100μm between pinholes. The MLM had 300 depth-graded layers of tungsten boron-carbide with an average thickness of 15Å each. Photons passing through different pinholes impinge on the MLM with slightly different angles of incidence which, in turn, flood the detector area with an large number of x-ray images each one characteristic of a slightly different photon energy range, i.e. an array of spectrally resolved images. The spatial resolution is $\Delta x \approx 10$μm, mainly limited by the pinhole size, and the spectral resolution power is $\lambda/\Delta\lambda \approx 150$. The array of spectrally resolved images recorded with MMI is rich in information, and can be processed to produce broad- and narrow-band x-ray images integrated over a range of photon energies or centered on line transitions of the tracer element, respectively [23]. Three identical MMI instruments are available for fielding in direct-drive implosions at OMEGA. More details about the instrument and data processing methods can be found in Refs. [19-23]. Other diagnostics fielded in the experiments included a neutron time-of-flight detector comprising a scintillator counter connected to a photomultiplier for total neutron yield measurements [24], a nuclear temporal diagnostic (NTD) [25] to measure the fusion reaction-rate time-history of the implosion and bang time (i.e., peak of the burn rate time-history), and a gated filtered (10mil Be and 0.5mil Ti), broadband x-ray imager (LFC) equipped with 10μm pinholes, six MCP's and capable of recording up to six images per MCP for a total of up to 36 images [26]. The Ti filter employed in LFC significantly reduces image contributions from photon energies above the Ti K-edge at 4.966keV. Hence, LFC broad-band images are dominated by contributions in the photon energy range from approximately 3keV to 4.966keV.

### 3. X-ray imaging observations and discussion

The goal of these x-ray imaging observations is to provide insight into the zero-order hydrodynamic characteristics of PDD implosions at OMEGA as extracted from observations along multiple LOS. The data were recorded along both equatorial and polar LOS, thus providing information about implosion performance as seen parallel to the equatorial plane and along the polar axis of the laser drive. The equatorial plane is defined as the plane perpendicular to the polar axis that cuts the target through its equator. We emphasize that these are the first, self-emission observations along the polar LOS reported for OMEGA PDD implosions well after the laser is shut off during the deceleration phase.

The images discussed here represent intensity maps on the image plane, and each intensity point (i.e. pixel) can be interpreted as the formal integral of the radiation transport equation along a ray (i.e. chord) through the plasma source (i.e. the imploded target), namely [27],

$$I_\nu = \int_0^L \varepsilon_\nu(x) e^{-\tau_\nu(x)} dx \qquad (1)$$

where $v$ is the photon frequency associated with a photon of energy $hv$, $I_v$ the emergent intensity, $\varepsilon_v$ the emissivity, and the integral runs over the entire length $L$ of the chord in the plasma source. The optical depth $\tau_v(x)$ of a point along the ray at physical depth $x$ is given by,

$$\tau_v(x) = \int_x^L k_v(x')dx' \qquad (2)$$

Note that Eq. (1) assumes that there is no external radiation incident on the plasma source. This is a good approximation since the images reported in this paper are due to self-emission of the target late in the implosion and not due to the transmission of photons through the target from a source of backlit photons. Experimental images are integrals of the monochromatic intensity $I_v$ over the photon energy band $\Delta hv$ characteristic of the recorded or reconstructed image,

$$I_{\Delta hv} = \int_{\Delta hv} I_v dhv \qquad (3)$$

We note that, in general, the higher the plasma electron temperature the larger the emissivity. But the emissivity also increases with density. Regions of large $\varepsilon_v$ are thus associated with high temperature and density. Hence, the regions of the plasma source that contribute the most to $I_v$ are those with the largest $\varepsilon_v$ so long as the optical depth is small.

Figures 1, 2 and 3 display self-emission x-ray images of the implosion recorded by an MMI instrument along a LOS 11° off the equatorial plane for the final stages of the implosion of OMEGA shot 60893. The MMI array of spectrally resolved images displays the Heα $1s^2$-1s2p at 4750eV and Heβ $1s^2$-1s3p at 5582eV lines in He-like Ti, and the Lyα 1s-2p transition at 4973eV in H-like Ti. The narrow-band Heβ image has a bandwidth of 130eV covering the photon energy range from 5510eV to 5640eV. This range includes the Heβ transition and Li-like Ti satellite lines on the low energy side of the Heβ parent line. The broad-band image has a bandwidth of 1500eV covering the range from 4600eV to 6100eV. The time sequence represents snapshots taken at t=1.7ns, t=1.8ns and t=1.9ns. The bang time for this shot determined by the peak of the burn time-history recorded with the NTD diagnostic was 1.9ns, and the laser was turned off by t=1.2ns. Hence, these images are characteristic of the deceleration phase and bang times of the implosion when the core driven by shock and compression heating becomes a hot and dense plasma, and emits x-ray radiation in the keV-range of photon energy due to bound-bound, bound-free and free-free atomic processes. The "hamburger" or "double-bun" structure observed in Fig. 1 is oriented perpendicular to the polar axis and can be seen in the MMI array of spectrally resolved images, in the reconstructed narrow-band Heβ image, and in the broad-band image thus indicating that this image feature is representative of a zero-order characteristic of the hydrodynamic behavior of the PDD implosion. In Fig. 1 each lobe or bright region of the "double-bun" has an approximate size of 90μm by 30μm, and a separation between lobes of about 20μm. Later in time (see Fig. 2) the two lobes join each other producing a smaller, single emission region of approximately 70µm by 80µm, while at bang time (see Fig. 3) the image shows an even smaller emission region, that we interpret as the implosion core, of size 40µm by 60µm, elongated along a direction perpendicular to the polar drive axis. We note that at bang time the Ti Heβ line is very weak, thus only the broadband image is shown.

To compare these observations with hydrodynamic simulations, two-dimensional (2D) Hydra simulations were performed of the experiment assuming axial symmetry about the polar axis. Hydra is an arbitrary Lagrangian Eulerian radiation hydrodynamics code [28]. Hydra does not include a mix model. It does have an extensive laser ray-tracing package to model the laser coupling to the target. The simulations reported here used the experimental beam pointing, though not the individual beam powers, i.e. each beam had the same, averaged power and hence effects due to power imbalance were not taken into account in the calculations. Hydra's total laser energy was decreased by 20% to approximately account for cross beam energy transfer (CBET) [29,30] and other absorption effects not included in the simulations. This adjustment was also needed in order to improve the matching between simulation and experimental bang times. After this adjustment, the bang time in the simulation was 1.7ns while in the experiment was 1.9ns. The simulation output temperature and density spatial profiles were post-processed to produce synthetic x-ray images for comparison with the experimental images. To this end, a synthetic diagnostic tool that includes the instrumental spatial resolution of 10µm was used to produce synthetic narrow band images in the same experimental photon energy range.

In the simulation, bang time occurred at t=1.7ns and the three snapshots selected for post-processing and image reconstruction shown in Fig. 4 are approximately equivalent to those of the gated images displayed in Figs. 1, 2 and 3 for the experiment. The synthetic images show characteristics similar to the experimental images. The earliest in time image in Fig. 4 shows the "double-bun" structure observed in the data and it is of comparable size. The intermediate and late time post-processed images in Fig. 4 show how the two bright lobes gradually join each other, and produce a single bright region at around bang time elongated along a direction perpendicular to the polar drive axis, as observed in the experiment.

The simulation also shows that this structure is associated with a zero-order characteristic of the hydrodynamics. In this connection, Fig. 5 displays the earliest in time, post-processed image of Fig. 4 for broad-band emission, and coarse-grain and spectroscopic-quality calculations of the narrow-band images centered in the Ti Heβ line for a bandwidth of 130eV. All of them show the same qualitative structure, consistent with the experimental observation. More detailed analysis of the simulation results are being conducted to better understand the origin of the "double bun" structure and will be reported elsewhere.

The "double-bun" structure observation along the LOS 11° off the equatorial view has been systematically observed in many PDD DIME campaign experiments at OMEGA. To illustrate this point, Fig. 6 shows gated images from a different series of experiments: OMEGA shot 67069. In this shot, two MMI instruments viewed the implosion along

LOS 11° off the equatorial view, but with different orientations in space. The polar axis is indicated in the figures, and it is always perpendicular to the "double-bun" structure. To compare with the PDD implosion observations, Fig. 7 shows an image recorded in OMEGA shot 65036 where similar targets were driven with a spherically symmetric, 60-beam drive [31]. No "double-bun" structure was observed in these experiment along any LOS.

Another systematic observation of PDD OMEGA experiments was provided by the MMI instrument that viewed the implosion along a polar axis LOS. The result is displayed in Figs. 8 and 9 for OMEGA shots 67066, 67071 and 67072. In this case, five bright spots can be seen in the image as well as an elongation of the shape of the image along the direction of the projection of the target's stalk on the image plane. We note in passing that a connection between stalk and core shape asymmetry has also been observed in 60-beam spherically symmetric illumination implosions at OMEGA [32]. The pentagonal pattern can be noted both in the array of spectrally resolved images recorded with MMI as well as in the narrowband and broadband reconstructed images. For later times, the images become smaller and appear more rounded, and the pattern is no longer seen in the data. The pentagonal pattern is also characteristic of a zero-order hydrodynamic feature of the PDD implosions. These five bright spots correlate and can be interpreted as a laser imprint effect due to the illumination pattern of the PDD configuration at OMEGA (see Fig. 9). More importantly, they reveal a three-dimensional (3D) aspect of the OMEGA PDD implosion associated with a laser imprint effect that has not been previously observed. This feature, first noticed in the MMI data, was also seen in gated, filtered broadband images recorded by the LFC instrument viewing the implosion along a polar LOS (see Fig. 10). Although the spectrally resolved image data recorded with MMI are more revealing, the LFC data did provide an independent confirmation of the petal pattern seen in the MMI images. Furthermore, the large time span covered by LFC also confirmed that this pattern is characteristic of the implosion core during the deceleration phase, and can last for up to 0.3 ns during the final stage of the implosion.

Motivated by the pattern observed in the polar view LOS, and since 2D Hydra simulations performed with axial symmetry about the polar axis cannot capture 3D effects, a preliminary 3D Hydra simulation of the experiment was performed in order to compare with the polar LOS observation. On the one hand, and as a consistency check, equatorial view synthetic images obtained from the 3D simulation were compared with those from the 2D simulation. This comparison showed good consistency between 2D and 3D synthetic images viewed along the equatorial LOS. On the other hand, preliminary results from polar view synthetic images obtained from post-processing the 3D simulations are consistent with the experimental observation and will be reported elsewhere.

Equatorial and polar view self-emission images reveal an asymmetric implosion core elongated in the equatorial plane (i.e. perpendicular to the polar axis), and with a pentagonal pattern observed along the polar axis that correlates with the OMEGA PDD laser drive configuration.

## 4. Conclusions

We have discussed spatially, temporally, and spectrally resolved narrow- and broad-band x-ray images of PDD OMEGA implosions. The self-emission images were obtained late in the implosion during the deceleration phase and around bang time, using several MMI instruments fielded along quasi-orthogonal LOS, including equatorial and polar views. The instruments recorded images based on K-shell lines from a titanium tracer in the shell as well as broad-band emission. These observations constitute the first such data obtained for PDD implosions, and reveal features attributed to zero-order hydrodynamics and laser imprinting.

Since the image data were recorded late in the implosion (well after the laser was turned off) they provide information on the shape and size of the implosion core, and patterns in the image intensity distribution during the deceleration phase, in the few hundred picoseconds leading up to bang time. The key observations from these MMI self-emission images are: (1) systematic differences in the images between equatorial and polar LOS, (2) polar views reveal a consistent pentagonal, petal pattern that correlates with the PDD laser illumination pattern used on OMEGA, and (3) the equatorial views reveal a "hamburger" or "double-bun" spatial structure. All three observations reveal zero-order hydrodynamic features about the implosion core assembly in that they are observed in the array of spectrally resolved images, and in both reconstructed narrow- (i.e., at a Ti K-shell line energy) and broad-band images, i.e. they are independent of the specific photon energy range and the radiation emission process.

Simulations performed with Hydra, both 2D and 3D, showed that equatorial LOS observations show good consistency with synthetic images obtained from post-processing simulation output. This is important since one of the goals of PDD experiments on OMEGA is to validate hydrodynamic simulations of PDD implosions. Preliminary results of synthetic polar view images from post-processing 3D Hydra simulations are encouraging in that they show the pentagonal pattern observed in the data, and will be reported elsewhere. Ongoing and future work include further spectroscopic analysis of the MMI data to extract the spatial distribution of temperature and density, as well as information on the amount of shell mix into the fuel. Finally, the observations discussed here emphasize the importance of recording both equatorial and polar view images in order to gain a more complete picture of the performance of PDD experiments.

**Acknowledgements**

This work was supported by a LANL ICF Program Academic Collaboration under DOE contract no. DE-AC52-06NA25396. We thank members of the Laboratory for Laser Energetics and LANL's P-24 group who assisted with diagnostic setup and operation.

# References

[a]Present address: Los Alamos National Laboratory, Los Alamos, NM.
[b]Present address: Sandia National Laboratories, Albuquerque, NM.

**Figure Captions**

**Figure 1:** array of spectrally resolved gated (t=1.7ns) images recorded with MMI along a LOS 11° off the equatorial plane (TIM3) in OMEGA shot 60893 (left). Ti He$\beta$ narrow-band (center) and broad-band (BB, right) reconstructed images. The arrow represents the projection of the polar axis on the image plane. Images are independently color coded to facilitate visualization of structure details; dark red is brightest.

**Figure 2:** reconstructed images as in Fig. 1 but for t=1.8ns. The arrow represents the projection of the polar axis on the image plane. Images are independently color coded to facilitate visualization of structure details; dark red is brightest.

**Figure 3:** reconstructed images as in Fig. 1 but for t=1.9ns, the experimental bang time. The arrow represents the projection of the polar axis on the image plane.

**Figure 4:** equatorial view synthetic broad-band images obtained by post-processing the output from a 2D Hydra simulation of OMEGA shot 60893 for t=1.56ns (left), t=1.65ns (center), and t=1.75ns (right). The arrow represents the projection of the polar axis on the image plane. Note: 1jk=1x10$^7$Joules.

**Figure 5:** equatorial view synthetic broad-band (left), Heβ coarse-grain (center), and Heb spectroscopic quality (right) narrow-band images obtained by post-processing the output from a 2D Hydra simulation of OMEGA shot 60893 for t=1.56ns. The arrow represents the projection of the polar axis on the image plane Note: 1jk=1x10$^7$Joules.

**Figure 6:** broad-band reconstructed images from gated data recorded with two MMI instruments fielded along two different LOS (TIM3 left, and TIM5 right) each 11° off the equatorial plane, in OMEGA shot 67069. For each LOS, the arrow indicates the projection of the polar axis on the image plane. Images are independently color coded to facilitate visualization of structure details; dark red is brightest.

**Figure 7:** broad-band reconstructed image from gated data recorded with MMI in OMEGA shot 65036.

**Figure 8:** array of spectrally resolved gated images recorded with MMI along the polar (TIM4) LOS in OMEGA shot 67066 (left). Ti Heβ narrow-band (center) and broad-band (BB, right) reconstructed images. The dashed arrow indicates the projection of the target stalk on the image plane. Images are independently color coded to facilitate visualization of structure details; dark red is brightest.

**Figure 9:** gated broad-band images reconstructed from MMI data recorded along the polar LOS (TIM4) for OMEGA shots 67071 and 67072. The dashed arrow indicates the projection of the target stalk on the image plane. Left and center images are independently color coded to facilitate visualization of structure details; dark red is brightest. Schematic illustration of the polar view of laser beams computed with VISRAD [33] (right).

**Figure 10:** gated, filtered broad-band image recorded with LFC along the polar LOS (TIM6) in shot 67071.

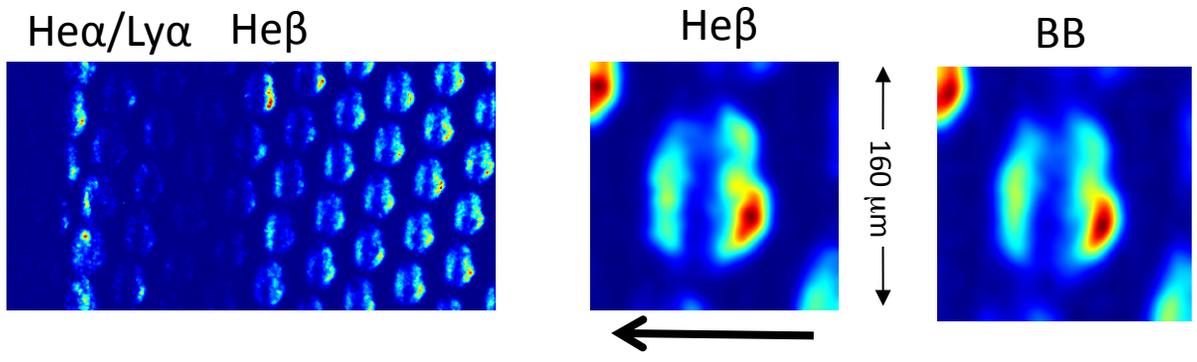

Figure 1

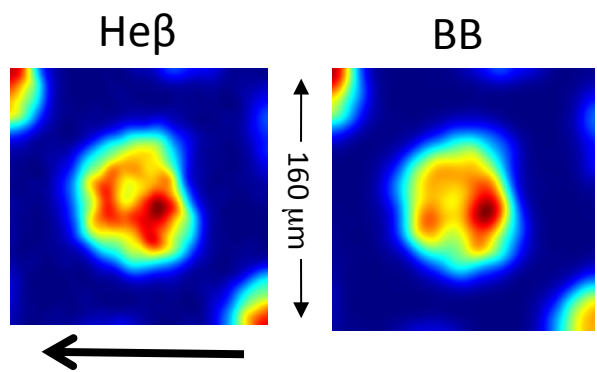

Figure 2

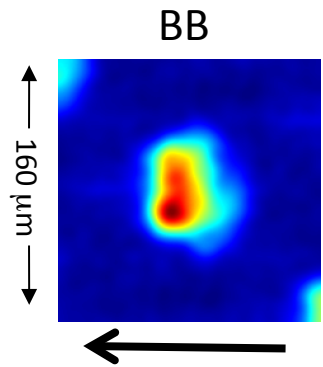

BB

160 μm

Figure 3

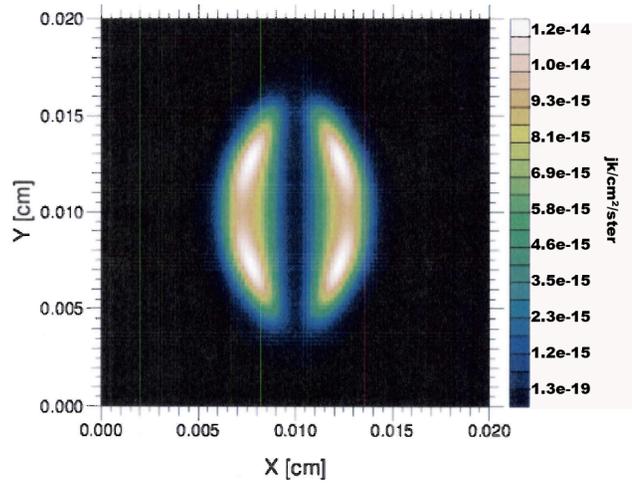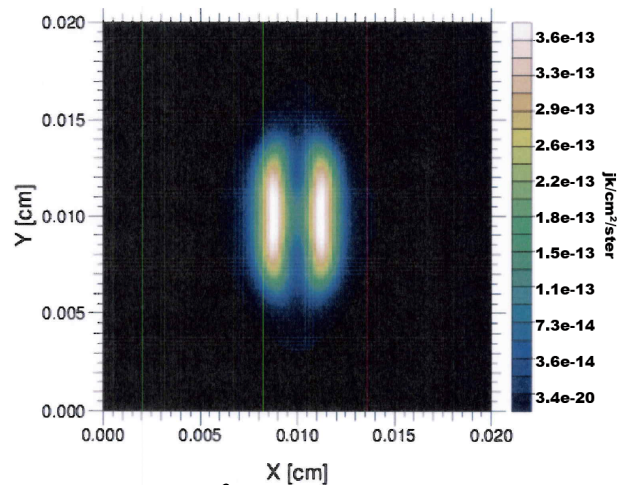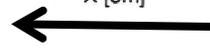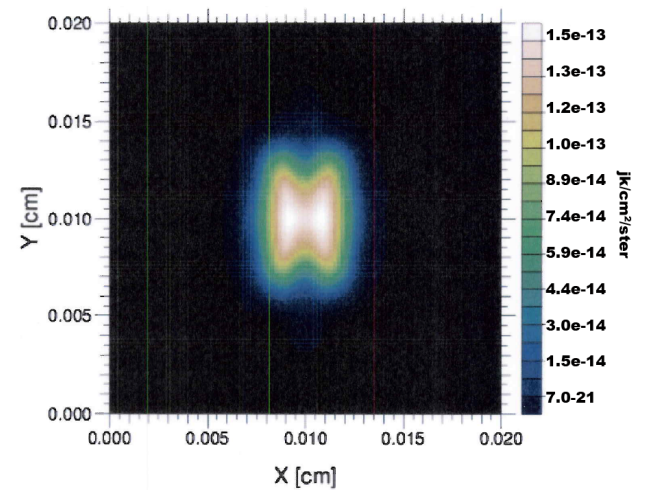

Figure 4

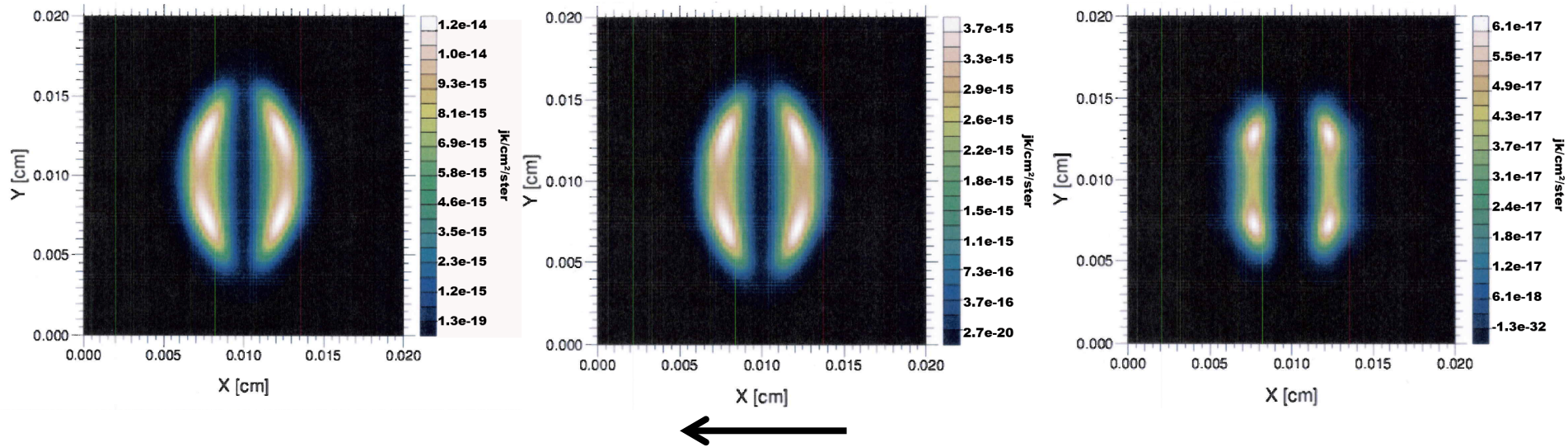

Figure 5

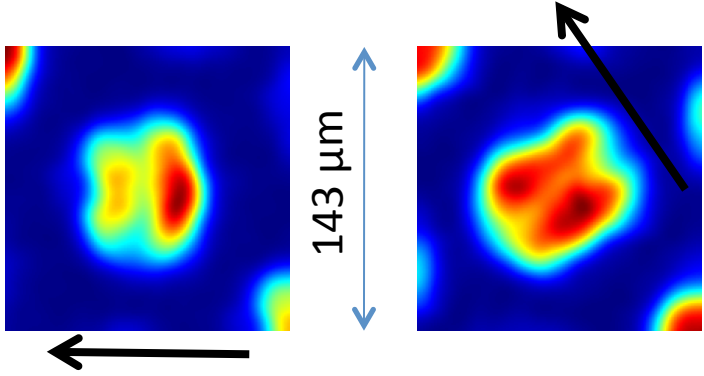

Figure 6

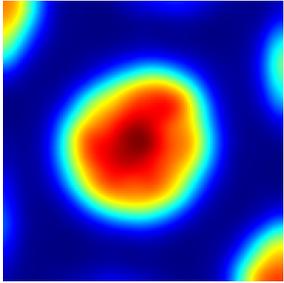

Figure 7

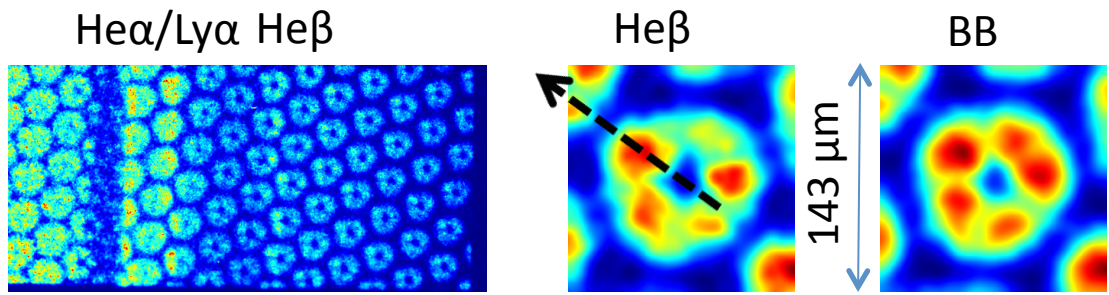

Figure 8

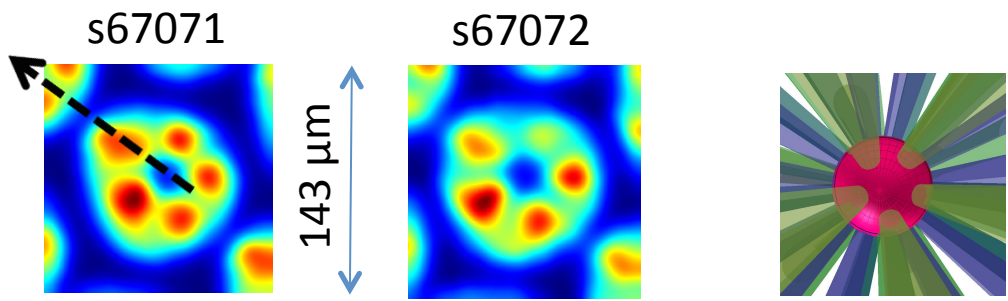

Figure 9

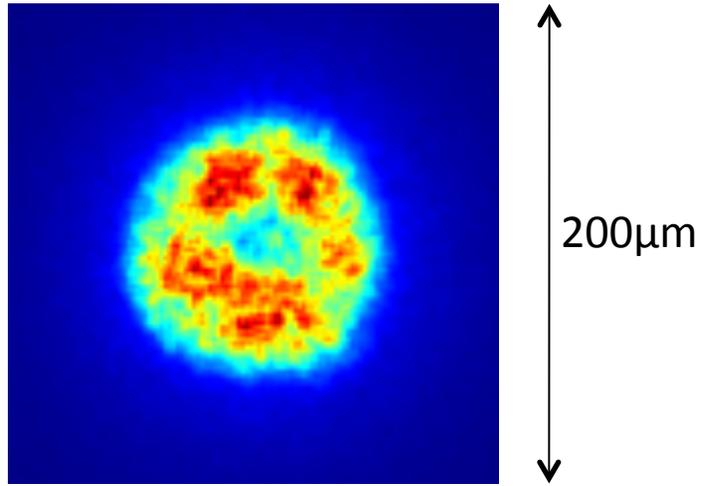

Figure 10